\documentclass[conference]{IEEEtran}
\usepackage{comment}
\usepackage{multicol}
\usepackage{blindtext}
\usepackage{amssymb}
\usepackage{subfigure}
\usepackage{amsmath}
\usepackage{amsfonts}
\usepackage{mathrsfs}
\usepackage{engord}
\usepackage[dvips]{graphicx}
\usepackage{bbm}
\usepackage{threeparttable}
\usepackage{booktabs}
\usepackage{epsfig}
\usepackage{color}
\usepackage{graphicx}
\usepackage{multirow}
\usepackage{cite}
\usepackage{epstopdf}
\usepackage{amsthm}

\usepackage{framed}
\usepackage{url}

\usepackage[ruled,norelsize]{algorithm2e}

\makeatletter
\newcommand{\removelatexerror}{\let\@latex@error\@gobble}
\makeatother
\begin{document}

\newtheorem{thm}{Theorem}
\newtheorem{prop}{Proposition}
\newtheorem{reproof}{Proof}
\newtheorem{lem}{Lemma}
\newtheorem{defn}{Definition}
\newtheorem{ex}{Example}
\newtheorem{cor}{Corollary}
\newtheorem{prn}{Principle}
\newtheorem{case}{Case}
%
\title{Algebraic Properties of Wyner Common Information Solution under Graphical Constraints}

\author{\IEEEauthorblockN{Md Mahmudul Hasan,
Shuangqing Wei, Ali Moharrer}}
\maketitle
\footnotetext[1]{Md M Hasan,   S. Wei and A. Moharrer  are with the school of Electrical Engineering and Computer Science, Louisiana State University, Baton Rouge, LA 70803, USA (Email: mhasa15@lsu.edu, swei@lsu.edu, alimoharrer@gmail.com). }



\begin{abstract}
The Constrained Minimum Determinant Factor
Analysis (CMDFA) setting was motivated by Wyner's common information problem where we seek a latent representation of a given Gaussian vector distribution with the minimum mutual information under certain generative constraints. In this paper, we explore the algebraic structures of the solution space of the CMDFA, when the underlying covariance matrix $\Sigma_x$ has an additional latent graphical constraint, namely, a latent star topology. 
In particular, sufficient and necessary conditions in terms of the relationships between edge weights of the star graph  have been found. Under such conditions and constraints, we have shown that the CMDFA problem has either a rank one solution or a rank $n-1$ solution where $n$ is the dimension of the observable vector.  Numerical results are provided to demonstrate the difference between the optimal mutual information and that derived under a naive star constraint.
\end{abstract}
\section{INTRODUCTION}
Factor Analysis (FA) is a commonly used tool in multivariate statistics to represent the correlation structure of a set of observables in terms of significantly smaller number of variables called ``latent factors". With the growing use in data mining, high dimensional data  analytics, factor analysis has already become a prolific area of research \cite{chen2017structured}\cite{bertsimas2017certifiably}. Classical Factor Analysis models seek to decompose the correlation matrix of an $n$-dimensional  random vector ${\bf X} \in {\mathcal R}^n$, $\Sigma_x  $, as the sum of a diagonal matrix $ D $ and a Gramian matrix $ \Sigma_{x}-D $. 

The literature that approached Factor Analysis can be classified in three major categories. Firstly, algebraic approaches \cite{albert1944matrices} and  \cite{drton2007algebraic}, where the principal aim was to give a characterization of the vanishing ideal of the set of symmetric $ n\times n $ matrices that decompose as the sum of a diagonal matrix and a low rank matrix, did not offer scalable algorithms for higher dimensional statistics. Secondly, Factor Analysis  via heuristic local optimization techniques, often based on the expectation maximization algorithm, were computationally tractable  but offered no provable performance guarantees. The third and final type are the  convex optimization based methods such as Constrained Minimum Trace Factor Analysis (CMTFA) \cite{bentler1980inequalities} \cite{hasan2018latent} and CMDFA \cite{moharrer2017algebraic}.  The motivation behind CMDFA comes from Wyner's common information $ C(X_{1},X_{2}) $ which characterizes  the minimum amount of common randomness needed to approximate the joint density between a pair of random variables $ X_{1} $  and $ X_{2} $ to be $C(X_{1},X_{2})= \min_{\underset{X_{1}-Y-X_{2}}{P_{Y}} } I(X_{1},X_{2}; Y) $, where $ I(X_{1},X_{2}; Y) $ is the mutual information between $ X_{1} $, $ X_{2} $ and $ Y $, $ X_{1}-Y-X_{2} $ indicates the conditional independence between $ X_{1} $ and $ X_{2} $ given $ Y $, and the joint density function is sought to esnure such conditional independence as well as the given joint density of $ X_{1} $  and $ X_{2} $. Since the Factor Analysis of the  Gaussian random vector $ \vec{X} $  can be modelled as $\vec{X}=A\vec{Y}+ \vec{Z}$,
where $A_{n\times k}$ is a real matrix, $ \vec{Y}_{k\times 1}, k<n $ is the vector of independent latent variables   and $\vec{Z}_{n\times 1}$ is a Gaussian vector of zero mean and covariance matrix $\Sigma_z = D$. Hence we have, 
$I(\vec{X};\vec{Y})=h(\vec{X})-h(\vec{X}|\vec{Y})=h(\vec{X})-h(\vec{Z})$
where $ I(\vec{X};\vec{Y}) $ is the mutual information between $ \vec{X} $ and $ \vec{Y} $, $ h(\vec{X}), h(\vec{Z})  $ are differential entropies of $ \vec{X}$ and $ \vec{Z} $ and $ h(\vec{X}|\vec{Y}) $ is the differential entropy of $ \vec{X} $ given $ \vec{Y} $.  Hence characterizing the common information between $ \vec{X} $ and $ \vec{Y} $ \cite{xu2016lossy}\cite{wyner1975common}\cite{satpathy2015gaussian}  would be $ \min_{A,\Sigma_{z}} I(\vec{X};\vec{Y}) $ which is an equivalent problem to $ \max_{\Sigma_{z}} h(\vec{Z})$ hence equivalent to $ \min_{\Sigma_{z}} -\log |\Sigma_{z}| $. 

The scope of this paper is limited to analysing the solution space of CMDFA and recovering the underlying graphical structures.  It is important to remark that our work is not concerned about the algorithm side of the optimization technique, rather our focus is to characterize and find insights about their solution space. Moharrer and Wei \cite{moharrer2017algebraic} derived CMDFA from a broader class of convex optimization problem and established  a relationship between the outcome of these optimization techniques and  common information problem \cite{wyner1975common}.  We find the explicit conditions under which the CMDFA solution of $ \Sigma_{x} $ recoves a star structure. Since star may not always be the optimum solution, we have also shown the existence and uniqueness of a rank $ n-1 $ CMDFA solution of $ \Sigma_{x} $ which is the only other possible solution. We have shown analyticaly the optimality of the non-star solution over star under certain circumstances from common information point of view, and at the end presened some numerical data to support our claims. 
\section{Definitions and Notations}
Let $ \vec{b} $ be a real $ n $ dimensional column vector  and $ A $ be an $ n\times n $ matrix. As in literature in general we denote the $ i $th element $ \vec{b} $ as $ b_{i} $ and the $ (i,j) $th element of $ A $ as $ A_{i,j} $. Here we define all the vector operations and notations in terms of  $ \vec{b} $ and $ A $, that will carry their meaning on other vectors and  matrices throughout this paper unless stated otherwise. 

 Let $ \vec{a}_{i,*} $ and $  \vec{a}_{*,i}  $ denote the $i$th row and $ i $th column vector of matrix $ A $ respectively. Function $ \lambda_{min}(A) $ is defined to be the smallest eigen value of matrix $ A $. $ N(A) $ stands for the null space of matrix $ A $. 

Vectors $ \vec{1} $ and $ \vec{0} $ are the $ n $ dimensional column vectors with each element equal to $ 1 $ and $ 0 $ respectively. When we write $ \vec{b}\geq 0 $ we mean that each element of the vector $ b(i)\geq 0, 1\leq i \leq n $.  $ \vec{b}^{2} $ is the Hadamard product of vector $ \vec{b} $ with itself. $ ||\vec{b} ||$ denotes the $ L_{2} $ norm of vector $ \vec{b} $. 

Now we define two terms i.e. \textit{dominance} and \textit{non-dominance} of a vector which will repeatedly appear throughout the paper. When we talk about the dominance or non-dominance of any vector $ \vec{b} $ we assume that the elements of the vector are sorted in a way such that $ |b_{1}|\geq |b_{2}|\geq \dots \geq |b_{n}|$. We call vector $ \vec{b} $ dominant and $ b_{1} $ the dominant element if for the above sorted vector $ |b_{1}|>\sum_{j\neq 1} |b_{j}| $  holds.  Otherwise $ \vec{b} $ is non-dominant.
\section{Formulation of the Problem} 
First of all we define the real column vector $ \vec{\alpha} $ as $\vec{\alpha} = [\alpha_1, \dots, \alpha_n]' \in {\mathcal R}^n$ where  $ 0< |\alpha_{j}|< 1 $, $ j= 1,2, \dots ,n $ and 
\begin{align}\label{order_alpha}
|\alpha_{1}|\geq |\alpha_{2}| \geq \dots \geq |\alpha_{n}|  
\end{align}
Let us consider a star structured population covariance matrix $ \Sigma_{x} $ having all the diagonal comptonents $ 1 $  as given by equation \eqref{sigmax}.
 \begin{equation}\label{sigmax}
\Sigma_{x}=\begin{bmatrix}
1&\alpha_{1}\alpha_{2}&\dots&\alpha_{1}\alpha_{n}\\
\alpha_{2}\alpha_{1}&1&\dots&\alpha_{2}\alpha_{n}\\
\vdots&\vdots&\ddots&\vdots\\
\alpha_{n}\alpha_{1}&\alpha_{n}\alpha_{2}&\dots&1\\
\end{bmatrix}
\end{equation}
The above covariance matrix could be produced by the graphical model given by equation \eqref{graphical model}. 
\begin{align}
& \begin{bmatrix}
X_{1}\\
\vdots\\
X_{n}
\end{bmatrix}
=\begin{bmatrix}
\alpha_{1}\\
\vdots\\
\alpha_{n}
\end{bmatrix}
Y
+
\begin{bmatrix}
Z_{1}\\
\vdots\\
Z_{n}
\end{bmatrix}\label{graphical model}\\
\Rightarrow &\vec{X}= \vec{\alpha}Y+\vec{Z}
\end{align}
where
\begin{itemize}
\item $ \{X_{1}, ..., X_{n}\}$  are conditionally independent Gaussian random variables given $ Y $, forming the jointly Gaussian random vector $ \vec{X}\sim \mathcal{N}(0,\Sigma_{x}) $ where $ Y\sim \mathcal{N}(0,1) $.
\item $ \{Z_{1}, ..., Z_{n} \}$ are independent Gausian random varables with $Z_{j}\sim \mathcal{N}(0,1-\alpha_{j}^{2})\quad 1\leq j \leq n $ forming the Gaussian random vector $ \vec{Z} $.
\end{itemize}
The above graphical model assumes the conditional independence among the observables given the latent variable given by \eqref{conditional independence} giving rise to a star topology.
\begin{align}\label{conditional independence}
p(X_{1}X_{2},\dots, X_{n}|Y)=\Pi_{i=1}^{n}p(X_{i}|Y)
\end{align}
CMDFA aims to minimize the mutual information between the observable Gaussian random vector $\vec{X}$ and the latent ones $\vec{Y}$. It is thus to seek joint distribution between the latent and observable ones such that the differential entropy $h(X|Y)$ is maximized. Under the joint Gaussian distribution, it is the same as seeking factorization of $\Sigma_x$ such that the determinant of the matrix $D$ is minimized as in equation \eqref{decomposition} under the constraint that both $ (\Sigma_{x}-D) $ and $ D $ are Gramian matrices.
\begin{align}\label{decomposition}
\Sigma_{x}=(\Sigma_{x}-D)+D
\end{align}
$ \Sigma_{x} $ being produced by the model in \eqref{graphical model} would equivalently mean  \eqref{decomposition} having a rank $ 1 $ solution i.e. $ \Sigma_{x}-D $ being $ \Sigma_{t,ND} $ given by equation \eqref{sigmatnd}.
\begin{align}\label{sigmatnd}
\Sigma_{t,ND}=\begin{bmatrix}
\alpha_{1}^{2}&\alpha_{1}\alpha_{2}&\dots&\alpha_{1}\alpha_{n}\\
\alpha_{2}\alpha_{1}&\alpha_{2}^{2}&\dots&\alpha_{2}\alpha_{n}\\
\vdots&\vdots&\ddots&\vdots\\
\alpha_{n}\alpha_{1}&\alpha_{n}\alpha_{2}&\dots&\alpha_{n}^{2}\\
\end{bmatrix}
\end{align}
But CMDFA solution of $ \Sigma_{x} $ may not always be rank $ 1 $, indicating star may not always be the optimum solution from common information point of view. It remains to be seen if  CMDFA solution to $ \Sigma_{x} $ recovers the graphical model given by \eqref{graphical model}. Also to be investigated is the exact solution to CMDFA if it fails to recover the underlying star topology. In the rest of the paper, we will present both sufficient and necessary conditions under which the rank of the optimal $\Sigma_{x} -D$ and the values of $D$'s entries are determined. 

\section{Solutions to CMDFA}
In this section we present the detailed analysis  of the CMDFA solution space of $ \Sigma_{x} $. We defne the real column vector $ \vec{\theta} \in {\mathcal R}^n$ as $\vec{\theta}=[\theta_{1},\dots, \theta_{n} ]'$ where $ \theta_{i}=\frac{|\alpha_{i}|}{\sqrt{1-\alpha_{i}^{2}}}, 1\leq i\leq n $.
 
As we can see, each elements in $ \vec{\theta} $ is equal to  the square root of the signal to noise ratio ($ \sqrt{\text{SNR}} $) of the corresponding element of vector $ \vec{\alpha} $. The following order of the elements of  $ \vec{\theta} $  is a necessary consequence of our assumption in \eqref{order_alpha},
\begin{align}\label{order_theta}
\theta_{1}\geq \theta_{2} \geq \dots \geq \theta_{n}
\end{align}
As we metioned  before, we are interested to find out if CMDFA low rank decomposition of $ \Sigma_{x} $ produces a rank $ 1 $ matrix. Next we analyse the solution space of CMDFA and find explicit conditions for both when the solution is rank $ 1 $ and when it is not. To start the proceedings we state Theorem \ref{sufficient and necessary cmdfa} given in \cite{moharrer2017algebraic} that gives
 the necessary and sufficient condition for $ D^{*}$ to be the CMDFA solution of the decomposition given in \eqref{decomposition}. 
\begin{thm}\label{sufficient and necessary cmdfa}
The matrix  $ D^{*}$ is the CMDFA solution of $ \Sigma_{x} $ if and only if $ \lambda_{min}( \Sigma_{x} -D^{*})=0 $, and there exists $ n\times k $ matrix $ T $ such that  $\vec{t}_{*,i}\in \mathcal{N}(\Sigma_{x}-D^{*}), 1\leq i \leq k$ and  $||\vec{t}_{i,*}||^{2}=\frac{1}{1-\alpha_{i}^{2}}, 1\leq i \leq n$.
\end{thm}

In the first of the two subsections of this section, we find  the conditions under which CMDFA solution of $ \Sigma_{x} $  recovers the model given by \eqref{graphical model} or equivalently speaking, find condtions under which CMDFA solution of $ \Sigma_{x} $ is the rank $ 1 $ matrix given by \eqref{sigmatnd}.   In the other subsection, we show the detailed analysis on the existance and uniqueness of the CMDFA solution of $ \Sigma_{x} $, when the solution is not a rank $ 1 $ matrix.
\subsection{CMDFA Non-dominant Case}
Here we analyse the conditions under which the CMDFA solution of $ \Sigma_{x} $ recovers a star structure. Lemma \ref{lem1} sets the groundwork for the Theorem to follow. The Lemma also has a geometric interpretaion that enriches our overall understanding of the CMDFA non-dominant case. 
\begin{lem}\label{lem1}
There exists $ n\times r $  matrix $ T $  such that $ \vec{t}_{*,i}\in N(\Sigma_{t,ND}), \quad 1\leq i \leq r $  and $ ||\vec{t}_{j,*}||^{2}=\frac{1}{1-\alpha_{j}^{2}}, \quad 1\leq j \leq n $ if and only if vector $ \vec{\theta} $  is non-dominant i.e.,
\begin{align}\label{non dominance of theta}
\theta_{1}\leq \sum_{i=2}^{n}\theta_{i}
\end{align}
\end{lem}
\begin{proof}[\textbf{Proof of Lemma \ref{lem1}:}]
Let $ \vec{t}_{i,*}, 1\leq i \leq n $ be the $ i $th row vector of the matrix $ T $ and $ \vec{0} $ denote the zero column vector. We need,
\begin{align}
&\Sigma_{t,ND}T=\vec{0}\notag\\
\Rightarrow &\vec{\alpha}T=\vec{0}\notag\\
\Rightarrow &\sum_{i=1}^{n}\alpha_{i}\vec{t}_{i,*}=\vec{0}\notag\\
\Rightarrow &\alpha_{1}\vec{t}_{1,*}=-\sum_{i=2}^{n}\alpha_{i}\vec{t}_{i,*}\notag\\
\Rightarrow &||\alpha_{1}\vec{t}_{1,*}||^{2}=||-\sum_{i=2}^{n}\alpha_{i}\vec{t}_{i,*}||^{2}\notag\\
\Rightarrow &||\alpha_{1}\vec{t}_{1,*}||^{2}\leq\sum_{i=2}^{n}||\alpha_{i}\vec{t}_{i,*}||^{2}\label{geo}
\end{align}
Equation \eqref{geo} has a beautiful geometric interpretation. The length of each vector  $ \alpha_{i}\vec{t}_{i,*} $ can be written as,
\begin{align}
& ||\alpha_{i}\vec{t}_{i,*}||^{2}=\alpha_{i}^{2}||\vec{t}_{i,*}||^{2}|| \alpha_{i}^2\sum_{j=1}^{n}t_{ij}^{2}=\frac{\alpha_{i}^{2}}{1-\alpha_{i}^{2}}, 1\leq i \leq n
\end{align}
\begin{figure}
\centering
\includegraphics[scale=0.5]{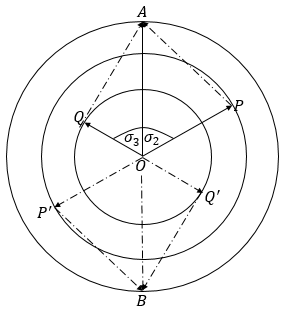}
\caption{Vectors on the surface of cocentric spheres. (dimension $ n=3 $)}
\label{fig3sphere}
\end{figure}
Hence, each vector $ \alpha_{i}\vec{t}_{i,*} $ is a point on the surface of an $ n $ dimensional sphere of radius $\frac{|\alpha_{i}|}{\sqrt{1-\alpha_{i}^{2}}}  $. Equation \eqref{geo} dictates that, for the matrix $ T $ to be in the null space of $ \Sigma_{t,ND} $ the biggest of those  spheres can not have a radius greater than the sum of the other radiuses. This leads us to the condition of non-dominance. Using \eqref{geo},
\begin{align}
&|\alpha_{1}| ||\vec{t}_{1,*}||\leq\sum_{i=2}^{n}|\alpha_{i}| ||\vec{t}_{i,*}||\notag\\
\Rightarrow & \frac{|\alpha_{1}|}{\sqrt{1-\alpha_{1}^{2}}}\leq \sum_{i=2}^{n}\frac{|\alpha_{i}|}{\sqrt{1-\alpha_{i}^{2}}}\notag\\
\Rightarrow & \theta_{1}\leq \sum_{i=2}^{n}\theta_{i}\notag
\end{align} 
\end{proof}
For further clarification, we refer to the  co-centric spheres (assuming $ n=3 $) in Figure \ref{fig3sphere}. Let $ ||\vec{OA}||=\theta_{1} $, $ ||\vec{OP}||=\theta_{2} $, $ ||\vec{OQ}||=\theta_{3} $. If $ \theta_{1}>\theta_{2}+\theta_{3} $, it is impossible to find any vector on the outer most sphere that can be expressed as the vector sum of vectors $ \vec{OP} $ and $ \vec{OQ} $. On the other hand if $ \theta_{1}\leq\theta_{2}+\theta_{3} $ proper selection of angles $ \sigma_{2} $ and $ \sigma_{3} $ can always ensure $\vec{OA}  $ be a vector sum of $ \vec{OP} $ and $ \vec{OQ} $ or equivalently ensure the orthogonality given by $ \vec{OB} +\vec{OP}+\vec{OQ}=\vec{0}$.

Having proved Lemma \ref{lem1} we are now well equipped to state and prove the statement of Theorem \ref{th3} that has the main result of this subsection.
\begin{thm}\label{th3}
CMDFA solution of $ \Sigma_{x} $ is $ \Sigma_{t,ND} $ if and only if $ \vec{\theta} $ is non-dominant. 
\end{thm}
The theorem states that the CMDFA solution to a star connected network is a star itself,  if and only if there is no dominant element in the vector $ \vec{\theta} $.
\begin{proof}[\textbf{Proof of Theorem \ref{th3}:}]
Now  we refer back to the necessary and sufficient condition for CMDFA solution at the begining of this section given by Theorem \ref{sufficient and necessary cmdfa}. Since, $ \Sigma_{t,ND} $ in rank $ 1 $, its minimum eigenvalue is $ 0 $. To complete the proof of Theorem \ref{th3}, we only need to show the existance of matrix $ T$ such that the column vectors of $ T$ are in the null space of $ \Sigma_{t,ND} $ and the  $ L_{2} $-norm square of the $ i$th row of $ T $ is $ \frac{1}{1-\alpha_{i}^{2}}, 1\leq i \leq n $. 

Lemma \ref{lem1} has already shown that, for the existence of such $ T $ non-dominance given by equation \eqref{non dominance of theta} is a necessary condition. Next we show, by constructing such a $ T $ matrix under the assumption of non-dominance of $ \vec{\theta} $, that non-dominace is also a sufficient condition . And that should complete the proof of Theorem \ref{th3}. 

It is straightforward to find  the following  basis vectors for the null space of  $ \Sigma_{t,ND} $,

\begin{align}
v_{1}=\begin{bmatrix}
-\frac{\alpha_{2}}{\alpha_{1}}\\
1\\
0\\
\vdots\\
0
\end{bmatrix}, v_{2}=\begin{bmatrix}
-\frac{\alpha_{3}}{\alpha_{1}}\\
0\\
1\\
\vdots\\
0
\end{bmatrix},\dots, \quad v_{n-1}=\begin{bmatrix}
-\frac{\alpha_{n}}{\alpha_{1}}\\
0\\
0\\
\vdots\\
1
\end{bmatrix}
\end{align}

We define matrix $ V $ so that its columns span the null space of $ \Sigma_{t,ND} $,
\begin{align}\label{v}
&V\notag\\
&= \begin{bmatrix}
-\frac{\alpha_{2}}{\alpha_{1}}&\dots&-\frac{\alpha_{n}}{\alpha_{1}}&-\left(c_{2}\frac{\alpha_{2}}{\alpha_{1}}+\dots+c_{n}\frac{\alpha_{n}}{\alpha_{1}}\right)\\
1&\dots&0&c_{2}\\
0&\dots&0&c_{3}\\
\vdots&\ddots&\vdots&\vdots\\
0&\dots&1&c_{n}
\end{bmatrix}
\end{align} 
where $ c_{i}=\frac{\widetilde{c}_{i}}{\sqrt{1-\alpha_{i}^{2}}}, \quad i=2,\dots, n $  and $ \widetilde{c}_{i}\in \{1,-1\} $.

The  columns of $ V $ span the null space of $ \Sigma_{t,ND} $. To construct our desired matrix $ T $, under the assumption of non-dominance of $ \vec{\theta} $, it will suffice for us to find a diagonal matrix $ B_{n \times n} $ such that the following holds. 
\begin{align}\label{t}
T_{n \times n}= V_{n \times n}\cdot B_{n \times n}
\end{align}
where the $ L_{2} $-norm square of the $ i $th row of $ T $ is $ \frac{1}{1-\alpha_{i}^{2}}  $. Using \eqref{t}, 
\begin{align}\label{tt}
TT'= VBB'V'=V \beta V'
\end{align}
We require the diagonal  matrix $ \beta $ to have only non-negative entries. Based on the conditions imposed on the matrix $ T $, we have the following $ n $ equations,

\begin{align}\label{d1}
&\frac{\alpha_{2}^{2}}{\alpha_{1}^{2}}\beta_{11}+\frac{\alpha_{3}^{2}}{\alpha_{1}^{2}}\beta_{22}+\dots+\frac{\alpha_{n}^{2}}{\alpha_{1}^{2}}\beta_{n-1,n-1}+\notag\\
&\left(c_{2}\frac{\alpha_{2}}{\alpha_{1}}+c_{3}\frac{\alpha_{3}}{\alpha_{1}}+\dots+c_{n}\frac{\alpha_{n}}{\alpha_{1}}\right)^{2}\beta_{nn}=\frac{1}{1-\alpha_{1}^{2}}
\end{align}

\begin{align}\label{d2}
\beta_{ii}+c_{i+1}^{2}\beta_{nn}=\frac{1}{1-\alpha_{i+1}^{2}}, \quad i=1,\dots,n-1
\end{align}

Solving, \eqref{d1} with the help of \eqref{d2} we get, 
\begin{align}
\beta_{nn}=
&\frac{\frac{\alpha_{1}^{2}}{1-\alpha_{1}^{2}}-\frac{\alpha_{2}^{2}}{1-\alpha_{2}^{2}}-\dots-\frac{\alpha_{n}^{2}}{1-\alpha_{n}^{2}}}{\sum_{i\neq j, i\neq 1, j\neq 1}c_{i}c_{j}\alpha_{i}\alpha_{j}}
\end{align}
It is straightforward to see that, to ensure all the $ \beta_{•ii} , 1\leq i \leq n$ are non-negative, we need $ \beta_{nn}\leq 1 $. We select $ \widetilde{c}_{i}, 2\leq i \leq n $ such that, 
\begin{align}
c_{i}\alpha_{i}=\frac{\widetilde{c}_{i}\alpha_{i}}{\sqrt{1-\alpha_{i}^{2}}}=\theta_{i}, \quad i=2,\dots, n
\end{align}
Under such selection of $ \widetilde{c}_{i}, 2\leq i \leq n $, $ \beta_{nn} $ becomes,
\begin{align}
\beta_{nn}=\frac{\theta_{1}^{2}-\theta_{2}^{2}-\dots-\theta_{n}^{2}}{\sum_{i\neq j, i\neq 1, j\neq 1}\theta_{i}\theta_{j}}
\end{align}
Now, using the non-dominance assumption given in \eqref{non dominance of theta}, we have
\begin{align}
&\theta_{1}^{2}\leq \left(\sum_{i=2}^{n}\theta_{i}\right)^{2}\notag\\
\Rightarrow &\frac{\theta_{1}^{2}-\sum_{i=2}^{n}\theta_{i}^2}{\sum_{i\neq j, i\neq 1, j\neq 1}\theta_{i}\theta_{j}}\leq 1\label{bc}\\
\Rightarrow & \beta_{nn}\leq 1
\end{align}
Which means non-dominance of vector $ \vec{\theta} $ is a sufficient condition to construct the kind of $ T $ matrix we are looking for. That completes the proof of Theorem \ref{th3}.
\end{proof}
\subsubsection*{Boundary Case}
It is obvious that, there might be numerous ways to construct the matrix $ T $ that satisfisfy the requirements set by Theorem \ref{sufficient and necessary cmdfa}.  Because of the special way we constructed the matrix $ T $ the rank of $ T $ under the non-dominant case is $ n-1 $ except for a very special boundary case.  Under the boundary case i.e. when the inequality \eqref{non dominance of theta} holds for equality, when the rank of $ T $ is always $ 1 $ irrespective of the way we construct $T $ .  For any given $ n $, it is straightforward to see from equation \eqref{bc} that,  for  $ \theta_{1}= \sum_{i=2}^{n}\theta_{i} $ we have $ \beta_{nn}=1 $. Plugging $ \beta_{nn}=1 $ in equation \eqref{d2} gives us $ \beta_{ii}=0, 2\leq i \leq n$. Equations \eqref{t} and \eqref{tt} suggest that, such a $ \beta $ matrix will produce a rank $ 1 $ matrix $ T $. This very special case is explained by the next Lemma. 
\begin{lem}\label{lem2}
When the non-dominance condition given in \eqref{non dominance of theta} holds for equality,
any $ n\times r $ matrix $ T $ such that 
$ \vec{t}_{*,i}\in N(\Sigma_{t,ND}), \quad 1\leq i \leq r $  and $ ||\vec{t}_{j,*}||^{2}=1, \quad 1\leq j \leq n $ has to be a rank $ 1 $ matrix.
\end{lem}
\begin{proof}[\textbf{Proof of Lemma \ref{lem2}:}]
Using the orthogonality between $ \Sigma_{t,ND} $ and its null space matrix $ T $,
\begin{align}\label{ortho}
\sum_{i=1}^{n}\alpha_{i}\vec{t}_{i,*}=\vec{0}
\end{align}
Equation \eqref{ortho} implies the following two things:
\begin{align}
&||\alpha_{1}\vec{t}_{1,*}||=||\sum_{i=2}^{n}\alpha_{i}\vec{t}_{i,*}||\label{imp1}\\
&\alpha_{1}\vec{t}_{1,*}  =-\sum_{i=2}^{n}\alpha_{i}\vec{t}_{i,*}\label{imp2}
\end{align}
Using the triangular inequality,
\begin{align}
||\sum_{i=2}^{n}\alpha_{i}\vec{t}_{i,*}||\leq \sum_{i=2}^{n}||\alpha_{i}\vec{t}_{i,*}||
\end{align}
If all the $ \alpha_{i}\vec{t}_{i,*}, 2\leq i \leq n $ are not in the same direction, the the above inequality becomes
\begin{align}
||\sum_{i=2}^{n}\alpha_{i}\vec{t}_{i,*}||< \sum_{i=2}^{n}||\alpha_{i}\vec{t}_{i,*}||
\end{align}
Hence, under  the boundary condition i.e. $ \theta_{1}= \sum_{i=2}^{n}\theta_{i} $ , we have
\begin{align}
||\sum_{i=2}^{n}\alpha_{i}\vec{t}_{i,*}||< ||\alpha_{1}\vec{t}_{1,*}||\notag
\end{align}
Which violets \eqref{imp1}. That means to ensure $||\alpha_{1}\vec{t}_{1,*}||=||\sum_{i=2}^{n}\alpha_{i}\vec{t}_{i,*}||$, all of $ \alpha_{i}\vec{t}_{i,*}, 2\leq i \leq n $ have to be in the same direction. This along  the second implication of orthogonality given by equation \eqref{imp2}, makes  matrix $ T $ a rank $ 1 $ matrix.
\end{proof}

\subsection{Dominant Case}
Having proved that the non-dominance of vector $ \vec{\theta} $ is a sufficient and necessary condition for CMDFA solution of $ \Sigma_{x} $ to recover a star structure, we are left with only the dominance case now i.e.
\begin{align}\label{domcon}
\theta_{1}>\sum_{i=2}^{n}\theta_{i}
\end{align}
Under the above dominant condition we want to show the existence of a rank $ n-1 $ solution of $ \Sigma_{x} $. Any solution we find will be unique, because CMDFA is a special type of the broader class of convex optimization problem defined in \cite{della1982minimum}. We still have to satisfy the same sufficient and necessary condtion for the CMDFA solution, that we presented at the begining of this section. Like the non-dominant case, for the matrix $ D^{*} $ to be the CMDFA solution of $ \Sigma_{x} $ under the dominant case, the minimum eigen value of $ \Sigma_{x} - D^{*}$ has to be $ \lambda_{min}(D^{*})=0 $ and the $ L_{2} $-norm square of the $ i $th row of the  null space matrix $ T $ has to be $ \frac{1}{1-\alpha_{1}^{2}} $. The only difference with the non-dominant case is that, since our conjecture for the dominant case is an $n-1  $ rank solution, the null space matrix $ T $ will  always be rank $ 1 $ i.e. a column vector. Mathematically speaking, we need to show the existance of $ 0<a_{i}<1, 1\leq i \leq n $ such that the following orthogonality condition holds.

\begin{align}\label{bigger picture}
\begin{bmatrix}
a_{1}&\alpha_{1}\alpha_{2}&\alpha_{1}\alpha_{3}&\dots&\alpha_{1}\alpha_{n}\\
\alpha_{2}\alpha_{1}&a_{2}&\alpha_{2}\alpha_{3}&\dots&\alpha_{2}\alpha_{n}\\
\vdots&\vdots&\vdots&\ddots&\vdots\\
\alpha_{n}\alpha_{1}&\alpha_{n}\alpha_{2}&\alpha_{n}\alpha_{3}&\dots&a_{n}\\
\end{bmatrix}
\begin{bmatrix}
\frac{c_{1}}{\sqrt{1-a_{1}}}\\
\vdots\\
\vdots\\
\frac{c_{n}}{\sqrt{1-a_{n}}}
\end{bmatrix}=
\begin{bmatrix}
0\\
\vdots\\
\vdots\\
0
\end{bmatrix}
\end{align}
where $ c_{i}\in \{ -1,1\} $. Once we have such $ a_{i}, 1\leq i \leq n$ the $ i $th element of the CMDFA solution vector  $ \vec{d^{*}} $ under the dominant case will be $ 1-a_{i},  1\leq i \leq n$. The  above orthogonality relationship gives us the following $ n  $ equations. 
\begin{align}\label{es}
&\frac{a_{i}c_{i}}{\sqrt{1-a_{i}}}+\sum_{j\neq i}\frac{\alpha_{i}\alpha_{j}c_{j}}{\sqrt{1-a_{j}}}=0, 1\leq i \leq n
\end{align}
Let $ (i) $ denote the $ i $th equation given by \eqref{es}. Using the linear combination $ \alpha_{i+1}\times (i)-\alpha_{i}\times (i+1), 1\leq i \leq n  $ gives us the following $ n-1 $ equations.
\begin{align}\label{a1}
\alpha_{i+1}c_{i}\eta_{i}-\alpha_{i}c_{i+1}\eta_{i+1}=0, \qquad 1\leq i \leq n-1
\end{align}
where
\begin{align}\label{eta}
\eta_{i}=\frac{a_{i}-\alpha_{i}^{2}}{\sqrt{1-a_{i}}}, \quad 1\leq i \leq n
\end{align}
Equation \eqref{a1} implies that for some ratio $ \mu$ we can write the following, 
\begin{align}\label{cb}
\begin{bmatrix}
c_{1}\eta_{1}\\
\vdots\\
c_{n}\eta_{n}
\end{bmatrix}=\mu\begin{bmatrix}
\alpha_{1}\\
\vdots\\
\alpha_{n}
\end{bmatrix}
\end{align}
Now plugging the expressions from \eqref{eta} and \eqref{cb} in any of the $ n $ equations given by \eqref{es} we get,
\begin{align}\label{main}
 \sum_{i=1}^{n}\frac{1}{1-\frac{a_{i}}{\alpha_{i}^{2}}}=1
\end{align}
It will suffice for us to prove the existence of $0<a_{i}<1, \quad 1\leq i \leq n$ such that \eqref{main} holds. From the definition of $ \eta_{i} $ given in \eqref{eta} we see that, to find each $ a_{i}, \quad 1\leq i \leq n$ we need to solve the following second order polynomial.
\begin{align}\label{poly}
a_{i}^{2}+a_{i}\alpha_{i}^{2}(\mu^{2}-2)+\alpha_{i}^{2}(\alpha_{i}^{2}-\mu^{2})=0, \quad 1\leq i \leq n 
\end{align}
If we solve equation \eqref{poly} for each $ a_{i} $ we will get a left root and a right root. Our initial conjecture is that the left root for $ a_{1} $ and right roots for $ a_{2},\dots, a_{n} $ that we get solving \eqref{poly} will give us $0<a_{i}<1, \quad 1\leq i \leq n$ that satisfy  \eqref{main}. If we can prove that our conjecture is true, then that should be the only possible solution to \eqref{main} because of the uniqueness of  solution to such convex optimization problems proved in \cite{della1982minimum}.
Plugging in the left root for $ a_{1} $, right roots for $ a_{2},\dots, a_{n} $ in \eqref{main} gives us the following equation.
\begin{align}\label{16}
1+\frac{1}{2}\sum_{i=1}^{n}\frac{\alpha_{i}^{2}}{1-\alpha_{i}^{2}}=&\frac{|\alpha_{1}|}{\sqrt{1-\alpha_{1}^{2}}}\sqrt{\frac{1}{4}\frac{\alpha_{1}^{2}}{1-\alpha_{1}^{2}}+\frac{1}{\mu^{2}}}\notag\\&-\sum_{i=2}^{n}\frac{|\alpha_{i}|}{\sqrt{1-\alpha_{i}^{2}}}\sqrt{\frac{1}{4}\frac{\alpha_{i}^{2}}{1-\alpha_{i}^{2}}+\frac{1}{\mu^{2}}}
\end{align}
We define
\begin{align}\label{xi}
X_{i}=\sqrt{\frac{1}{4}+\frac{\frac{1}{\mu^{2}}}{\frac{\alpha_{i}^{2}}{1-\alpha_{i}^{2}}}}=\sqrt{\frac{1}{4}+\frac{\frac{1}{\mu^{2}}}{\theta_{i}^{2}}}, \quad i=1,2,\dots, n
\end{align}
Under these newly defined $ X_{i} $s \eqref{16} becomes,
\begin{align}\label{plane}
 &\theta_{1}^{2}X_{1}-\sum_{i=2}^{n}\theta_{i}^{2}X_{i}=1+\frac{1}{2}\sum_{i=1}^{n}\theta_{i}^{2}
 \end{align}
 And using the definition of $ X_{i}, 1\leq i \leq n $ given in \eqref{xi}, we get the following cylinders of hyperbolas.
 \begin{align}\label{hyperbolas}
 &\theta_{1}^{2}X_{1}^{2}-\theta_{i}^{2}X_{i}^{2}=\frac{1}{4}(\theta_{1}^{2}-\theta_{i}^{2}), \quad 2\leq i \leq n
 \end{align}
 
 Equations given by \eqref{hyperbolas} imply that for each value of $ X_{1} $ we get a point $ [X_{1}, X_{2}, \dots ,X_{n} ],$ in the $ n $ dimensional space where each $ X_{i}, 2\leq i \leq n $ is a function of $ X_{1} $. For the range of values  of $ (\frac{1}{2} <X_{1}<\infty) $ all such points together produce an $ n $ dimensional space curve. If we project this space curve on any of the two dimensional $ (X_{1},X_{i}), 2\leq i \leq n $ planes we get a hyperbola. 

 Another important thing to note is that, each equation given by \eqref{hyperbolas} is a cylinder of hyperbolas originated from $ (X_{1},X_{i}) $ plane and projected onto $ n $ dimensional space. Each point in the space curve represents an intersection points of all $ n-1 $ cylinders of hyperbolas originated from $ (X_{1},X_{i}), 2\leq i \leq n $ planes.
 
 At this point our revised goal is to show the existence of  a point in the space curve that satisfies  equation \eqref{plane} under the dominance condition given by \eqref{domcon}. Becasue of the way we defined $ X_{i} $s $ 1\leq i \leq n $ the solution must satisfy the condition $ X_{i}>\frac{1}{2}, 1\leq i \leq n$. Theorem \ref{thm2} states the main result of this subsection. 
 
 \begin{thm}\label{thm2}
 There exists an intersection point among the plane given by \eqref{plane} and the $ n-1 $ cylinders of hypberbolas given by \eqref{hyperbolas}, that satisfies $ X_{i}>\frac{1}{2}, 1\leq i \leq n $.
 \end{thm}
 
 Proving the above Theorem would mean that, there exists $0<a_{i}<1, \quad 1\leq i \leq n$ such that \eqref{main} holds, which in turn would mean the existance of an $ n-1 $ rank CMDFA solution under the dominance of vector $ \vec{\theta}$. And as we mentioned already, the uniqueness of such solution is guaranteed.  
 
 \begin{proof}[\textbf{Proof of Theorem \ref{thm2}:}]
 Let us define the function $ G(.) $ of $ X_{1} $ as the inner product between the vectors $ [X_{1}, \dots ,X_{n}] $ and $ [\theta_{1}^{2}, \dots ,\theta_{n}^{2}]' $ where each $ X_{i}, 1\leq i\leq n$ is a  function of $ X_{1} $. Which means,
\begin{align}\label{G}
&G(X_{1})=\theta_{1}^{2}X_{1}-\sum_{i=2}^{n}\theta_{i}^{2}X_{i}(X_{1})
\end{align}

So, our revised goal becomes to find the existence of such $ X_{1}>\frac{1}{2} $ for which the function of $ G(X_{1}) $ becomes $ G(X_{1})=1+\frac{1}{2}\sum_{i=1}^{n}\theta_{i}^{2} $. And to achieve that goal some functional analysis of $ G(X_{1}) $ that we present next are of paramount importance.
\begin{figure}
\centering
\includegraphics[scale=0.5]{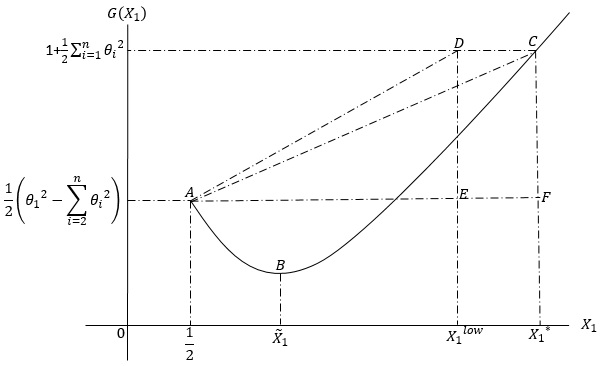}
\caption{Trend of the function $ G(X_{1} )$ against $ X_{1} $}
\label{figgx}
\end{figure}
\begin{figure}
\centering
\includegraphics[scale=0.5]{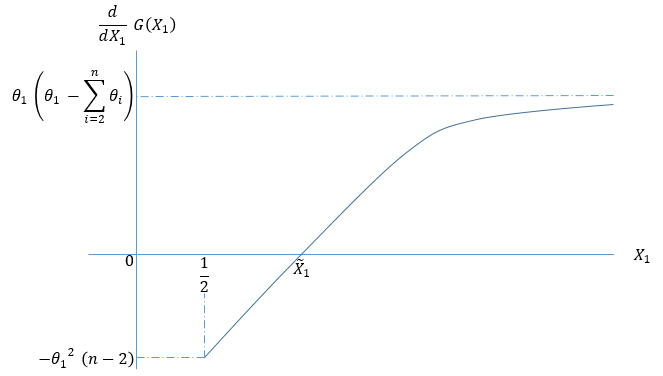}
\caption{Trend of the function $ \frac{d}{dX_{1}}G(X_{1} )$ against $ X_{1} $}
\label{figdgx}
\end{figure}

Equation \eqref{hyperbolas} dictates that each  $ X_{i}(X_{1}), 2\leq i \leq n$ is a concave function of $ X_{1}>\frac{1}{2} $. Which makes $ G(X_{1}) $ given by \eqref{G} a convex function of $ X_{1} $ as the sum of convex functions of $ X_{1} $. 
Using \eqref{hyperbolas} and \eqref{G} we get,
\begin{align}\label{halfG}
G\left(\frac{1}{2}\right)=\frac{1}{•2}\left(\theta_{1}^{2} -\sum_{i=2}^{n}\theta_{i}^{2}\right)
\end{align}
Using \eqref{xi} we get,
\begin{align}\label{dx}
\frac{dX_{i}(X_{1})}{dX_{1}}&=\frac{\frac{dX_{i}(X_{1})}{d\nu}}{\frac{dX_{1}}{d\nu}}=
\frac{\frac{1}{•2X_{i}(X_{1})}\frac{1}{\theta_{i}^{2}}}{\frac{1}{•2X_{1}}\frac{1}{\theta_{1}^{2}}}=\frac{\theta_{1}^{2}X_{1}}{\theta_{i}^{2}X_{i}(X_{1})}
\end{align}
where, $ \nu=\frac{1}{\lambda^{2}} $. Using \eqref{G} and \eqref{dx},
\begin{align}
&\frac{dG(X_{1})}{dX_{1}}=\theta_{1}^{2}\left[ 1-\sum_{i=2}^{n}\frac{X_{1}}{X_{i}(X_{1})}\right]\label{ratio}
\end{align}
Hence,
\begin{align}\label{half}
\Rightarrow &\left.\frac{dG(X_{1})}{dX_{1}}\right\vert_{X_{1}=\frac{1}{2}}=-\theta_{1}^{2}(n-2)
\end{align}

which is a negative value. We define $ \hat{X}_{1} $ such that, 
\begin{align}\label{minG}
\Rightarrow &\left.\frac{dG(X_{1})}{dX_{1}}\right\vert_{X_{1}=\hat{X}_{1}}=0
\end{align}
Figures \ref{figgx} and \ref{figdgx} illustrate our findings from the above functional analysis. 
As each $ X_{i}(X_{1}), 2\leq i \leq n$ is an increasing function of $ X_{1} $ the ratios $\frac{X_{1}}{X_{i}(X_{1})}, 2\leq i \leq n $ are decreasing functions of $ X_{1} $. Hence equation \eqref{ratio} suggests that $ \frac{dG(X_{1})}{dX_{1}} $ is an increasing function of $ X_{1} $. Given that knowledge, equations \eqref{half} and \eqref{minG} considered together imply $ \hat{X}_{1}>\frac{1}{2} $ as seen in  Figures \ref{figgx} and \ref{figdgx}. One important to remark is that we see the function $ \frac{dG(X_{1})}{dX_{1}} $ gets saturated gradually and is upperbounded by a value. This is because the ratio $\frac{X_{i}(X_{1})}{X_{1}}, 2\leq i \leq n $ is the slope of  hyperbola in $ X_{1}-X_{i} $ plane which is upper bounded by $ \frac{\theta_{1}}{\theta_{i}} $ which is the slope of the asyptote in the respective plane. Plugging these individual upperbounds in \eqref{dx} we get the dotted upper bound in Figure \ref{figdgx}. 

We can argue, as we refer to Figure \ref{figgx}, since $ G $ is a convex function of $ X_{1} $, it must be an increasing function for the values  $ X_{1}> \hat{X}_{1} $. Equations \eqref{minG} and \eqref{halfG} imply that the convex function value $G(\hat{X}_{1})<G\left(\frac{1}{2} \right)<1+\frac{1}{2}\sum_{i=1}^{n}\theta_{i}^{2}  $. Hence, there must exist $ X_{1}^{*}>\hat{X}_{1}>\frac{1}{2}  $ such that $ G(X_{1}^{*})=1+\frac{1}{2}\sum_{i=1}^{n}\theta_{i}^{2} $.
 \end{proof}
 From Theorem \ref{thm2} and its proof we know that   the solution $ X_{1}^{*} $ produces a corresponding $ n  $ dimensional point $ \vec{X}^{*}=[X_{1}^{*}, \dots , X_{n}^{*}]' $ in the space curve which is the intersection point among hyperbolic cylinders  and the plane given by \eqref{hyperbolas} and \eqref{plane} respectively.  If we reflect on the bigger picture, $ X_{1}^{*} $ plugged in \eqref{xi} will produce a value of $ \mu $ which in turn plugged in \eqref{poly} will give us a set of $ a_{i}, 1\leq i \leq n $ that satisfies \eqref{bigger picture}. 
 \subsubsection{Bounds of the Solution}
 Here we find a lowerbound and an upperbound to $ X_{1}^{*} $.
 
 \textbf{Upperbound to $ X_{1}^{*} $:} 
   It is easy to derive that the $ i $th hyperbolic cylinder given by \eqref{hyperbolas} has the following corresponding equation of the cylinder asymptotes  (the ones passing through the origin and the first quadrant of the respective plane). 
 \begin{align}\label{asymptotes}
  X_{i}=\frac{\theta_{1}}{•\theta_{i}}X_{1}, \quad 2\leq i \leq n
  \end{align}
  Solving \eqref{asymptotes} and \eqref{plane} together we get a value of $ X_{1} $ which we denote  as $ X_{1}^{up} $ given by \eqref{xup}, 
  \begin{align}\label{xup}
  X_{1}^{up}=\frac{1+\frac{1}{•2}\sum_{i=1}^{n}\theta_{i}^{2}}{•\theta_{1}(\theta_{1}-\sum_{j=2}^{n}\theta_{j})}
  \end{align}
 Substituting  $ X_{1} $ in \eqref{asymptotes} by $ X_{1}^{up} $ gives us a vector $ \vec{X}^{up}=[X_{1}^{up},\dots,X_{n}^{up}]' $ in the $ n $ dimensional space, which is the intersection of the cylinders of asymptotes in \eqref{asymptotes} and the plane in \eqref{plane}. 
  
  \begin{lem}\label{upperbound lemma}
  The intersection point among the plane in \eqref{plane} and the hyperbolic cylinders in \eqref{hyperbolas} is upperbounded by  the intersection point among the same plane and asymptotes of the respective hyperbolic cylinders given by \eqref{asymptotes},
  \end{lem}
  The prooof of Lemma \ref{upperbound lemma} is given in Appendix A. According the statement of this Lemma $ \vec{X}^{up}>\vec{X}^{*} $. Which immediately suggests that $ X_{1}^{up} $ given by \eqref{xup} is an upperbound on $ X_{1}^{*} $. 
 
  \textbf{Lowerbound to $ X_{1}^{*} $:}   We see in Figure \ref{figgx} that the average slope of the curve $ ABC $ is captured by the slope of the line $ AC $. We assume that the dashed line $ AD $ in Figure \ref{figgx}  has slope $ \theta_{1}\left(\theta_{1}- \sum_{i=2}^{n}\theta_{i}\right) $ i.e. the upperbound of $ \frac{dGX_{1}}{dX_{1}} $ given in Figure \ref{figdgx}.  Figure \ref{figdgx} suggests that, the slope  at each point of the curve   $ ABC $ is strictly less than the slope of  $ AD $ in Figure \ref{figgx}, hence the slope of $ AC $ must be less than the slope of $ AD $. Now considering triangles $ \triangle ADE $ and $ \triangle ACF $ in Figure \ref{figgx} we have, 
  \begin{align}
  &\frac{DE}{AE}>\frac{CF}{AF}\notag\\
  \Rightarrow&\frac{DE}{AE}>\frac{DE}{AF}\notag\\
  \Rightarrow& AF>AE\notag\\
  \therefore& X_{1}^{*}>X_{1}^{*}
  \end{align}
 Which suggests $ X_{1}^{low} $ a lowerbound of the actual  $ X_{1}^{*} $. Next we find the expression for $ X_{1}^{low} $ using the geometry in Figure \ref{figgx}.
  
  \begin{align}
 X_{1}^{low}&=\frac{1}{•2}+AE\notag\\
 &=\frac{1}{2}+\frac{DE}{\frac{DE}{AE}}\notag\\
 &=\frac{1}{2}+\frac{1+\frac{1}{2}\sum_{i=2}^{n}\theta_{i}^{2}}{\theta_{1}\left(\theta_{1} -\sum_{i=2}^{n}\theta_{i}\right)}\label{xlow}
 \end{align}
 \section{Numerical Data}
 We motivated CMDFA in terms  common information which is a function of the minimum mutual information between the observables and the latent factors. It is a common practice to assume the star topology i.e the assumption that all the observables are mutually independent given a latent factor. Though star offers a sparce structure and smooth analysis, it may not be always the optimum solution. Next we show that assumption of star under CMDFA dominant case does not produce optimum outcome from common information point of view. We show that under the dominant case CMDFA solution provides lower mutual information between the observables and the latent variables that the star solution. Which in turn means lower common randomness required to produce the joint distribution between the observables and the latent variables and hence lower Wyner common information.  In summary, we are about to demonstrate the additional cost in using more information bits to synthesize $n$-dimensional Gaussian vector under a star topology, when we do not use the solution of CMDFA, under the dominant case. 
 
 As mentioned before, each of $ X_{1}^{low} $ and $ X_{1}^{up} $  will produce a corresponding $ \mu $ from equaion \eqref{xi} and a set $ a_{i}, 1\leq i \leq n $ or equivalently produce a matrix $ \Sigma_{z} $ that decomposes \eqref{decomposition}. Let $ X_{1}^{low} $ and $ X_{1}^{up} $ produce $\mu^{low} $ and $\mu^{up} $ from equaion \eqref{xi}, the corresponding sets $\{a_{i}^{low}\}_{i=1}^{n}  $ and $\{a_{i}^{up}\}_{i=1}^{n}  $ from \eqref{poly}, corresponding matrices  $ \Sigma_{z}^{low} $ and $ \Sigma_{z}^{up} $  that decompose \eqref{decomposition}, and $ I^{low} $, $ I^{up} $ be the corresponding mutual information between observed variables and the latent variables    respectively. Also let $ \Sigma_{z}^{star} $ be the solution to \eqref{decomposition} when the CMDFA solution is a star and  $ I^{star} $ be the corresponding mutual information between the observed variables and the latent factor .  
  Next Theorem analytically shows that each of $ I^{low} $, $ I^{up} $  produces better results than $ I^{star} $ considered from common information point of view. We present the comparative results with respect to  the varying magnitude of the dominance of $ \vec{\theta} $. Referring to equation \eqref{domcon}, we vary the dominance of vector $ \vec{\theta} $ by changing the value of the first element $ \theta_{1} $ while keeping other elements unchanged. 
 \begin{thm}\label{th7}
 All of $ I^{star}-I^{low} $, $ I^{star}-I^{up} $ and $ I^{up}-I^{low} $ are increasing functions of $ \theta_{1} $
 \end{thm}
 \begin{figure}
\centering
\includegraphics[scale=0.65]{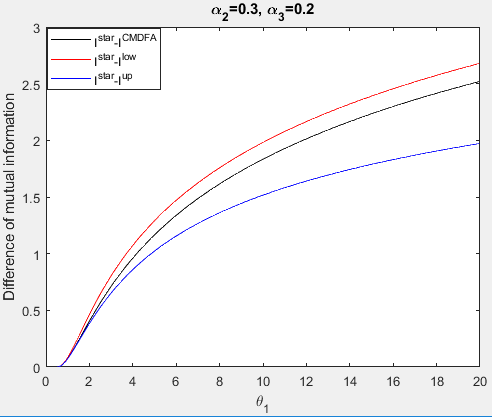}
\caption{Difference of mutual information against $ \theta_{1} $}
\label{fignumerical}
\end{figure}
The proof of Theorem \ref{th7} is given in Appendix B. Figure \ref{fignumerical} presents $ I^{star}-I^{low} $, $ I^{star}-I^{up} $ and $ I^{star}-I^{CMDFA} $ as functions of $ \theta_{1} $ for a particular case of $ n=3 $, where $ I^{CMDFA} $ is the mutual information between the observed variables and the latent ones corresponding to the numerically found solution $ X_{1}^{*} $. As we mentioned in the introduction that the primary motivation for this part of the work comes from common information, and the fact that in general people tend to assume a star topology to find common information, any value of mutual information less than $ I^{star}$ works to our advantage. $ I^{star}-I^{up} $ is an increasing function of $ \theta_{1} $ indicates that the lower bound of the advantage of CMDFA solution over star increases as vector $ \vec{\theta}$ becomes more and more dominant. We numerically calculated $I^{CMDFA} $ and the curve in Figure \ref{fignumerical} gives the actual advantage that CMDFA soution has over star under the dominance of $ \vec{\theta} $ whereas $ I^{star}-I^{low} $ gives an upperbound to the actual advantage of CMDFA over a star topology. The gap between $ I^{star}-I^{low} $ and $ I^{star}-I^{up} $ is gradually increasing indicating $ I^{up}-I^{low} $ is increasing with $ \theta_{1} $ which justifies the statement of Theorem \ref{th7}.
 \section{Conclusion}
In this paper we analyzed the solution spaces of   convex optimization algorithm  CMDFA.  We found conditions under which the solution is a star (rank $ 1 $) and proved the existence of a rank $ n-1 $ solution when the solution is not a star. Through analytical analysis followed by numerical data we showed that star is not always the optimum solution. We particularly demonstrated the additional cost in using more information bits to synthesize $n$-dim Gaussian vector under a star topology, when we do not use the solution of CMDFA, under the  dominant case. 
\appendices
\section{•}
\begin{proof}[\textbf{Proof of Lemma \ref{upperbound lemma}:}]
\begin{figure}
\centering
\includegraphics[scale=0.7]{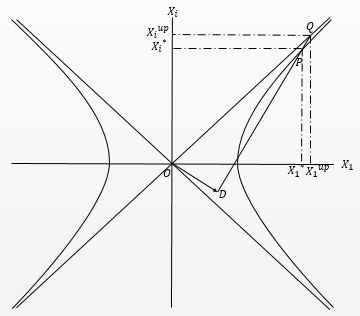}
\caption{Projection of the $ n $ dimensional intersection poin on $ X_{1}-X_{i} $ plane }
\label{fighyperbola}
\end{figure}
Let $ \vec{X}^{*}= [X_{1}^{*}, \dots , X_{n}^{*} ]'$ be the CMDFA solution vector i.e. the intersection point of \eqref{hyperbolas} and \eqref{plane}.  We refer to Figure \ref{fighyperbola}, the CMDFA dominant case solution vector $ \vec{X}^{*}= [X_{1}^{*}, \dots , X_{n}^{*} ]'$  has been projected on  the $ (X_{1},X_{i} )$ plane which is shifted in the direction of $ X_{j}, j\neq 1, j\neq i $ by $ X_{j}^{*} $. The procection of the $ n $ dimensional plane given by \eqref{plane} on this $ (X_{1},X_{i}) $ plane is given by the line $ DPQ $ which is at a perpendicular distance $ OD $ (because $ \angle ODP = 90^{\circ} $) from the origin. Here $ OD $ is the projection of the vector $ [\theta_{1}^{2}, -\theta_{2}^{2},- \dots , -\theta_{n}^{2}]' $ on $ (X_{1},X_{i}) $ plane, whose length we can calculate from equation \eqref{plane} as,
\begin{align}
OD=1+\frac{1}{2}\sum_{m=1}^{n}\theta_{m}^{2}-\sum_{j\neq 1, j\neq i}\theta_{j}^{2}X_{j}^{*}
\end{align}
Geometrically, we can see in Figure \ref{fighyperbola} that the  line $ OPQ $ which is the projection of the plane cuts the hyperbola at point $ P $ and the corresponding asymptote  at point $ Q $ in the first quadrant of the $ (X_{1},X_{i}) $ plane. It is obvious to notice that, because of the higher elevation and sharper slope of the asymptote compared to the hyperbola, point $ Q $ has higher  coordinate values than point $ P $ which is the projection of the CMDFA solution vector $ \vec{X}^{*} $ on $ (X_{1},X_{i})$ plane i.e. $X_{1}^{up}>X_{1}^{*}  $ and $X_{i}^{up}>X_{i}^{*}  $. 
The above conclusion holds true for any projection of $ \vec{X}^{*} $ on any  $(X_{1},X_{i}),  2\leq i \leq n $ plane. For example, the projection on $(X_{1},X_{2}) $ plane will give us $X_{1}^{up}>X_{1}^{*}  $ and $X_{2}^{up}>X_{2}^{*}  $.  Combining the outcome of all such projection for $ 2\leq i \leq n $ we can conclude that $ \vec{X}^{up}>\vec{X}^{*} $. Which algebraically means, the intersection point among the hyperbolic cylinders in \eqref{hyperbolas} and the plane in \eqref{plane} is upper bounded by the intersection point among the asymptotes of the respective hyperbolic cylinders given by \eqref{asymptotes} and the plane in \eqref{plane}.  
\end{proof}
\section{•}
\begin{proof}[Proof of Theorem \ref{th7}:]
Before we go into the business part of the proof we do some general preparatory groundwork. Using Equation \eqref{poly} and the fact that we used the right root of $ a_{1} $ we get,
 \begin{align}
 a_{1}&=\frac{1}{2}\left[\alpha_{1}^{2}(2-\mu^{2})- \sqrt{\alpha_{1}^{4}(2-\mu^{2})^{2}-4\alpha_{1}^{2}(\alpha_{1}^{2}-\mu^{2})} \right]\notag\\
\Rightarrow\frac{1-a_{1}}{1-\alpha_{1}^{2}} &=1+\frac{\alpha_{1}^{2}\mu^{2}}{•2-2\alpha_{1}^{2}}+\frac{1}{2}\frac{\alpha_{1}^{2}}{1-\alpha_{1}^{2}}\sqrt{•(2-\mu^{2})^{2}-4+\frac{4\mu^{2}}{•\alpha_{1}^{2}}}\notag\\
\Rightarrow\frac{1-a_{1}}{1-\alpha_{1}^{2}} &=1+\frac{\mu^{2}\theta_{1}^{2}}{2}+\frac{\theta_{1}^{2}}{•2}\sqrt{-4\mu^{2}+\mu^{4}+\frac{4\mu^{2}}{\alpha_{1}^{2}•}}\notag\\
\Rightarrow\frac{1-a_{1}}{1-\alpha_{1}^{2}} &=1+\frac{\mu^{2}\theta_{1}^{2}}{2}+\frac{\theta_{1}^{2}}{•2}\sqrt{\mu^{4}+\frac{4\mu^{2}}{\theta_{1}^{2}•}}\notag\\
\Rightarrow\frac{1-a_{1}}{1-\alpha_{1}^{2}} &=1+\frac{\mu^{2}\theta_{1}^{2}}{2}+\frac{\mu^{2}\theta_{1}^{2}}{•2}\sqrt{1+\frac{4}{\mu^{2}\theta_{1}^{2}•}}\notag\\
\Rightarrow\frac{1-a_{1}}{1-\alpha_{1}^{2}} &=1+\frac{\mu^{2}\theta_{1}^{2}}{2}+\frac{\mu^{2}\theta_{1}^{2}}{•2}\sqrt{•\frac{4}{\mu^{2}\theta_{1}^{2}•}}+\notag\\ &\frac{\mu^{2}\theta_{1}^{2}}{•2}\left( \sqrt{1+\frac{4}{\mu^{2}\theta_{1}^{2}•}}-\sqrt{•\frac{4}{\mu^{2}\theta_{1}^{2}•}}\right)\notag\\
\Rightarrow\frac{1-a_{1}}{1-\alpha_{1}^{2}} &=1+\frac{\mu^{2}\theta_{1}^{2}}{•2}+\sqrt{•\mu^{2}\theta_{1}^{2}}+\notag\\&\frac{\mu^{2}\theta_{1}^{2}}{•2}\left( \sqrt{1+\frac{4}{\mu^{2}\theta_{1}^{2}•}}-\sqrt{•\frac{4}{\lambda^{2}\mu_{1}^{2}•}}\right)\label{a1}
 \end{align}
 Similarly, since we are using the right roots for 
 $ a_{i}, 2\leq i \leq n $ we get, 
 \begin{align}\label{2ton}
 \Rightarrow\frac{1-a_{i}}{1-\alpha_{i}^{2}} &=1+\frac{\mu^{2}\theta_{i}^{2}}{2}+\frac{\mu^{2}\theta_{i}^{2}}{•2}\sqrt{•\frac{4}{\mu^{2}\theta_{i}^{2}•}}-\notag\\ &\frac{\mu^{2}\theta_{i}^{2}}{•2}\left( \sqrt{1+\frac{4}{\mu^{2}\theta_{i}^{2}•}}-\sqrt{•\frac{4}{\mu^{2}\theta_{i}^{2}•}}\right)\notag\\
 \end{align}
 Equations \eqref{xlow} and \eqref{xup} suggest that both $ X_{1}^{low} $ and $ X_{1}^{up} $ are decreasing functions of $ \theta_{1}^{2} $. And in turn  equation \eqref{xi} suggests that $ \mu \theta_{i}^{2}, 1\leq i \leq n $ are increasing functions of $ \theta_{1}^{2} $. Since $ \theta_{i} , 2\leq i \leq n$ are constants, the only thing changing in \eqref{2ton} is $\mu$. But $ \mu $ can not increase beyond $ \alpha_{1}^{2} $ because that would mean equation \eqref{main} does not have a solution. Hence, in order to increase $ \theta_{1} $ as we keep on increasing the value of $ \alpha_{1} $ and make it closer and closer to $ 1 $, the value of $ \mu $ also gets closer to $ 1 $. Thus with the increament of $\theta_{1}^{2}$ the parameters given by \eqref{2ton} asymptotically converge to  constants which we get plugging in $ \mu=1 $ i.e. for $ 2\leq i \leq n $
 \begin{align}
 \Rightarrow\frac{1-a_{i}}{1-\alpha_{i}^{2}} =1+2\theta_{i}+\frac{\theta_{i}^{2}}{2}\left(1-\sqrt{1+\frac{4}{•\theta_{i}^{2}}} \right),\quad 2\leq i \leq n 
 \end{align}
 Now that we have the groundwork done, we can proceed to prove the actual statement of the theorem. 
 \begin{align}
 I^{star}-I^{low}&=\frac{1}{•2}\log |\Sigma_{z}^{low}|-\frac{1}{•2}\log |\Sigma_{z}^{star}|\notag\\
 &=\frac{1}{2}\log\sum_{i=1}^{n}\frac{1-a_{i}^{low}}{1-\alpha_{i}^{2}}\notag\\
 &=\frac{1}{2}\log\frac{1-a_{1}^{low}}{1-\alpha_{1}^{2}}+\sum_{i=2}^{n}\frac{1}{2}\log\frac{1-a_{i}^{low}}{1-\alpha_{i}^{2}}\label{star-low}
 \end{align}
 $ \sum_{i=2}^{n}\frac{1-a_{i}^{low}}{1-\alpha_{i}^{2}} $ is asymptotically  a constant. Equation \eqref{xi} suggests $ \mu^{low} $ is an increasing function of $ \theta_{1} $ because $ X_{1}^{low} $ is a decreasing function of $ \theta_{1} $. Hence from \eqref{a1},
 \begin{align}
 \Rightarrow\frac{1-a_{1}^{low}}{1-\alpha_{1}^{2}} &=1+\frac{(\mu^{low})^{2}\theta_{1}^{2}}{•2}+\sqrt{•(\mu^{low})^{2}\theta_{1}^{2}}\notag\\
 &+\frac{(\mu^{low})^{2}\theta_{1}^{2}}{•2}\left( \sqrt{1+\frac{4}{(\mu^{low})^{2}\theta_{1}^{2}•}}-\sqrt{•\frac{4}{(\mu^{low})^{2}\theta_{1}^{2}•}}\right)
 \end{align}
 Which is an increasing function of $ \theta_{1} $. Hence from \eqref{star-low} we see that $ I^{star}-I^{low} $ is an increasing function of $ \theta_{1} $. Similarly,
 \begin{align}
 I^{star}-I^{up}=\frac{1}{2}\log\frac{1-a_{1}^{up}}{1-\alpha_{1}^{2}}+\sum_{i=2}^{n}\frac{1}{2}\log\frac{1-a_{i}^{up}}{1-\alpha_{i}^{2}}\label{star-up}
 \end{align}
 Like the previous case we can argue that, $ \sum_{i=2}^{n}\frac{1-a_{i}^{up}}{1-\alpha_{i}^{2}} $ is asymptotically  a constant. Equation \eqref{xi} suggests $ \mu^{up} $ is an increasing function of $ \theta_{1} $ because $ X_{1}^{up} $ is a decreasing function of $ \theta_{1} $. Hence $ \frac{1-a_{1}^{up}}{1-\alpha_{1}^{2}} $ and consequently $I^{star}-I^{up}  $ is an increasing function of $ \theta_{1} $. 
 
 Using equations \eqref{star-low} and \eqref{star-up}, 
 \begin{align}
 I^{up}-I^{low}=&\frac{1}{2}\log\frac{1-a_{1}^{low}}{1-\alpha_{1}^{2}}-\frac{1}{2}\log\frac{1-a_{1}^{up}}{1-\alpha_{1}^{2}}+\kappa\notag\\
 &=\frac{1}{2}\log\frac{1-a_{1}^{low}}{1-a_{1}^{up}}+\kappa
 \end{align}
 where $ \kappa $ is a constant. Since  $a_{1}^{low}  $  and $ a_{1}^{up} $ are increasing functions of $ \mu^{low}\theta_{1}^{2} $ and $ \mu^{up}\theta_{1}^{2} $ respectively, to show $ I^{up}-I^{low} $ is an increasing function of $ \theta_{1} $ we need to show $ \frac{\mu^{low}}{\mu^{up}} $ is an increasing function of $ \theta_{1} $. Equations \eqref{xlow} and \eqref{xup} suggest that $ \frac{X_{1}^{up}}{X_{1}^{low}} $ is an increasing function of $ \theta_{1} $, and in turn \eqref{xi} suggests $ \frac{\mu^{low}}{\mu^{up}} $  is an increasing function of $ \theta_{1} $. That completes the final part of the proof. 
 \end{proof}
\bibliography{ref}
\bibliographystyle{IEEEtran}

\end{document}